\documentclass[11pt,aps,prd,preprint]{revtex4}

\usepackage[dvips]{color}
\usepackage{epsfig}
\usepackage{amsmath}
\usepackage{graphicx}
\usepackage{subfigure}
\date{\today}

\usepackage[dvips]{color}
\usepackage{epsfig}
\usepackage{amsmath}
\usepackage{graphicx}
\usepackage{subfigure}
\date{\today}
\begin{document}
\title{Traversable wormholes in Finsler geometry under conformal motion}
\author{Manjunath Malligawad $^{1}$}
\email{manjunathmalligawad91@gmail.com}
\author{S. K. Narasimhamurthy $^{1}$}
\email{nmurthysk@gmail.com}
\author{Z. Nekouee $^{2}$}
\email{zohrehnekouee@gmail.com}
\author{Rajesh Kumar $^{3}$}
\email{rkmath09@gmail.com}
\author{Mallikarjun Y. Kumbar $^{4}$}
\email{mallikarjunykumbar@gmail.com}

\affiliation{ $^{1} $ \small{Department of PG Studies and Research in Mathematics,
		Kuvempu University\\ Jnana Sahyadri, Shankaraghatta - 577 451, Shivamogga, Karnataka, India.}}
\affiliation{ $^{2} $ \small{School of Physics, Damghan University, Damghan, 3671641167, Iran.}}
\affiliation{ $^{3} $ \small{Department of Mathematics and Statistics,\\  DDU Gorakhpur University, Gorakhpur, Uttaar Pradesh, India.}}
\affiliation{ $^{4} $ \small{Department of Mathematics,  Mahantswamy Arts, Science and Commerce College,\\  Haunsbhavi - 581 109, Haveri, Karnataka, India.}}

\begin{abstract}
This paper aims to investigate the possibility of physically achievable new wormhole solutions within the context of Finsler geometry by focusing on conformal motion. For this purpose, we study the barotropic linear equation of state (EoS). We scrutinized their geometric features within the EoS model, which includes baryonic and non-baryonic matter. Through derived field equations, shape functions adhering to critical criteria governed by the Finsler parameter $\gamma$ are explored. By evaluating energy conditions, violations are found, which indicate the potential presence of exotic matter necessary for traversable wormholes. Notably, violations of energy conditions signify the plausibility of traversable wormholes within the conformally transformed Finslerian space observed in both baryonic and non-baryonic scenarios. Additionally, anisotropy exploration uncovers repulsive geometric characteristics within these wormholes.
\end{abstract}
\maketitle
\newpage
\section{Introduction}\label{sec.1}
	Two distinct space-times can potentially be connected through a shortcut, forming what is known as a "wormhole", a hypothetical topological phenomenon in space-time. The term "wormhole" was initially coined by Fuller and Wheeler \cite{Fuller}, who credited Weyl \cite{Weyl} for the foundational idea of non-simply connected space-time. Einstein and Rosen later delved into the wormhole geometry, unveiling the Einstein-Rosen bridge and investigating non-singular coordinate patches within the Reissner-Nordstr$\ddot{o}$m and Schwarzschild solutions. It's essential to distinguish between these two fundamental concepts of wormholes: those arising from eternal black hole solutions, unsuitable for information transfer and inherently non-traversable, and those emerging from the non-simply connected nature of space-time proposed by Weyl, which predict electromagnetic field lines without a source, potentially serving as a classical communication channel. The inquiry into the feasibility of constructing a sustainable wormhole as a topological tunnel in space in alignment with the laws of physics has spurred significant curiosity among researchers. In their research paper, Einstein and Rosen \cite{Einstein} introduced theories regarding wormholes, initially proposing the existence of non-traversable wormholes with an event horizon that restricts the movement of observers. However, for the purpose of traversing between distinct universes, a traversable wormhole serves as the most straightforward means, devoid of an event horizon and singularity.

	In recent years, there has been significant fascination with traversable Lorentzian wormholes hypothetical narrow passages that potentially link two regions within the same universe or even different universes. The attention towards these structures intensified notably after Morris and Thorne's groundbreaking research \cite{Morris}. These wormholes, akin to 'shortcuts' in space-time, emerge from the implications of the Einstein field equations \cite{Morris,Visser1}, positioning them within the hierarchy of black holes and white holes.
	It is noted that most of the wormhole solutions have been devoted to the study of static configurations that must satisfy some specific properties in order to be traversable.
The primary feature of such a wormhole involves the presence of exotic matter encompassing the throat region, which stands as a crucial factor in maintaining this stable configuration and necessitates a deviation from the null energy condition \cite{Ida,Jamil}.
Deviations from this energy condition are a significant concern \cite{Hochberg}. Visser, cited in \cite{Visser1}, demonstrated the possibility of constructing wormhole space-time with minimal violations of the averaged null energy conditions. A. Banerjee et al. \cite{Banerjee} employed Conformally Symmetric methods to investigate traversable wormholes within f(R, T) gravity, while P. Bhar \cite{Bhar} explored the feasibility of sustaining static and spherically symmetric traversable wormhole structures by integrating conformal motion within Einstein's gravity framework. Hengfei Wu \cite{Hengfei} studied traversable phantom wormholes employing conformal symmetry within the context of $f(R, \phi, \xi$) gravity while utilizing the linear equation of state.  In recent studies, the exploration of wormhole models in $f(R, T)$ gravity has gained attention \cite{Bhar1}. Mustapa, in their work \cite{Mustafa4}, delved into novel wormhole solutions in teleparallel gravity, coupling minimal matter and incorporating conformal symmetry with non-zero Killing vectors.
Furthermore, Munsif  \cite{Munsif} investigated potential new wormhole solutions within the framework of $f(Q)$ gravity, leveraging the concept of conformal symmetry. Numerous researchers have introduced solutions for wormholes that rely on exotic forms of matter within both General Relativity (GR) and modified gravitational theories \cite{Wang,Lobo,Singh,Rahaman3}.
The exploration of diverse space-time properties within extended gravity theories has garnered significant interest among scientists lately. Notably, a predominant focus of scholarly articles involves the geometric facets of the field equations. Furthermore, recent empirical observations have prompted a quest for a more comprehensive gravitational theory beyond  GR. The progress in quantum gravity studies has led to numerous inquiries suggesting that GR might be encompassed within a broader framework of gravitational fields.
	
	The crucial rationale favoring the selection of Finsler geometry over GR stems from the fact that (i) Finsler geometry's reliance on both system dynamics and position obviates the necessity for quadratic constraints to confine the length element \cite{Bao}. The temporal interval perceived by an observer between two events equates to the spatial separation between two events unfolding along the observer's world line. (ii) This measurement is contingent upon the tangent bundle of a homogeneous function, leading to arcs influenced by both length and velocity. Top of the form (iii) Moreover, the present Finsler space, which demonstrates quadratic characteristics concerning $y_\theta$ and $y_{\phi}$, can be readily addressed by examining $F(x, y) = \sqrt{(g_{ij} y^i y^j)} $within the Riemannian manifold $(M, g_{ij}(x))$ the present Finsler space which is quadratic in terms of $y_\theta$ and $y_{\phi}$. Within Finsler geometry, the Weyl theorem elucidates that the projective and conformal attributes of a Finsler metric uniquely define its geometric properties. Consequently, the conformal aspects of Finsler metrics have captivated the interest of numerous researchers. B$\acute{a}$cs$\acute{o}$ and Cheng \cite{Bacso, Chen} have delineated specific conformal transformations that preserve the integrity of Riemann curvatures, Ricci curvatures, Landsberg curvatures, mean Landsberg curvatures, or S-curvatures. This advancement has spurred the identification of numerous Finsler geometry models across diverse physical applications, as evidenced by the scholarly works referenced in \cite{Vacaru,Vacaru1,Schreck}. Consequently, Finsler geometry has garnered considerable interest over recent decades due to its capacity to elucidate phenomena beyond the explanatory scope of Einstein's gravitational framework \cite{Rutz,Kostelecky,Kouretsis,Li3,Vacaru3}. This approach has led numerous researchers to unearth wormhole solutions within the realm of Finsler geometry \cite{Manjunatha,Singh,Paulb,Manju}.
	
	Within the realm of wormhole physics, the methodology employed offers a divergent path in tackling Einstein's field equations.
Here, the sequence involves formulating the space-time metric before deriving the stress-energy tensor components. Exploring a more scientific avenue for attaining precise solutions holds significance. One objective of this investigation involves examining wormhole solutions within Finsler geometry that accommodate conformal motion. Embracing the concept of conformal motion within Finslerian geometry delves into geometric transformations that maintain object shapes while disregarding alterations in size, holding profound implications. Within this context, we embrace an approach delineated in references like \cite{Maartens,Herrera}, presuming spherical symmetry and conformal symmetry in the space-time structure. The distinct characteristic of hereditary symmetry within symmetric configurations owes its existence to conformal Killing vectors (CKVs), offering valuable insights into spatial-temporal geometry. This study unveils a wormhole solution within the framework of Finslerian geometry, taking into account conformal motion and deriving the associated field equation. Moreover, we ascertain the conformal factor and shape function by exploring equations of state (EoS). Additionally, our discourse delves into a discussion concerning the potential violation of energy conditions exhibiting the stable wormhole model.

The paper is organized as follows: In section (\ref{sec.2}), we establish the field equation utilizing Finsler geometry. Section (\ref{sec.3}) outlines the development of a Finslerian wormhole that incorporates conformal motion. In section (\ref{sec.4}), we investigate particular solutions for the wormhole, considering barotropic linear EoS with baryonic and non-baryonic fluid. In subsection (\ref{subsec4.1}) and (\ref{subsec4.2}), we discuss two wormhole models with their physical viability, proper radial distance, radial velocity and energy conditions, and effect of anisotropy via graphical representation. Lastly, the article concludes in section (\ref{sec.5}), offering a summary and conclusions drawn from the research.

	\section{Metric formalism and basic equations of Finslerian gravity}\label{sec.2}
	The Finsler geometry rooted in the Finsler structure $F$ is defined as per the description in \cite{Bao} represented by the equation $F(x,\kappa y)=\kappa F(x,y)$ for all positive values of $\kappa$. Here, $x$ denotes the position within a four-dimensional smooth manifold denoted as $M$, and $y=dx/dt$ signifies velocity. The expression of the metric tensor $g_{\mu\nu}$ for Finsler geometry is
	\begin{equation}\label{Eq.2}
		g_{\mu\nu}=\frac{1}{2}\frac{\partial^2 F^2 (x, y)}{\partial y^\mu \partial y^\nu}.
	\end{equation}
	In Finsler space-time, the geodesic equation describes the paths of free particles moving under the influence of gravity and other related forces. The geodesic equation in Finsler space-time is given as:
	\begin{equation}\label{Eq.3}
		\frac{d^{2}x^{\mu}}{d\tau^{2}}+2G^{\mu}(x, y)=0,
	\end{equation}
	where $G^{\mu}$ denotes the geodesic spray coefficient, and it is stated  as
	\begin{equation}\label{Eq.4}
		G^{\mu}=\frac{1}{4}g^{\mu\nu}\left(\frac{\partial^{2}F^{2}}{\partial x^{\iota}\partial y^{\nu}}y^{\iota}-\frac{\partial F^{2}}{\partial x^{\nu}}\right).
	\end{equation}
	The Akbar-Zadeh \cite{Akbar-Zadeh} proposed equation for the Finslerian modified Ricci tensor, and it is written as
	\begin{equation}\label{Eq.5}
		Ric_{\mu\nu}=\frac{\partial^{2}\bigg(\frac{1}{2}F^2 Ric\bigg)}{\partial y^{\mu}\partial y^{\nu}},
	\end{equation}
	where $Ric$ is the Ricci scalar, which is the invariant value in Finsler space-time, and it is expressed in Finsler geometry as \cite{Bao},
	\begin{equation}\label{Eq.6}
		Ric\equiv R^\mu_\mu=\frac{1}{F^2}\left (2\frac{\partial G^\mu}{\partial x^\mu}-y^\iota \frac{\partial^2 G^\mu}{\partial x^\iota  \partial y^\mu}+2G^\iota  \frac{\partial^2 G^\mu}{\partial y^\iota  \partial y^\mu}-\frac{\partial G^\mu}{\partial y^\iota } \frac{\partial G^\iota }{\partial y^\mu}\right),
	\end{equation}
	\begin{equation}\label{Eq.7}
		R^\mu_\nu=\frac{1}{F^2}R^\mu_{\iota\nu\sigma}y^\iota y^\sigma,
	\end{equation}
	where $R^\mu_\nu$ depends mainly on Finsler structure $F$ and $R^\mu_{\iota\nu\sigma}$.\\
	
	Since almost all of the static vacuum solutions can be reduced to Schwarzschild's form according to Birkhoff's theorem, Finsler geometry extends the conventional Riemannian geometry by introducing a new metric structure that allows for the consideration of more general geometric spaces, so we will consider that the Finsler structure $F$ is of the form,
	\begin{equation}\label{Eq.8}
		F^2=e^{v(r)}y^t y^t-e^{\lambda(r)} y^r y^r-r^2 \bar{F}^2(\theta, \phi, y^\theta, y^\theta).
	\end{equation}
	Here we have the expressions $v(r)=2a(r)$ and $e^\lambda=\left(1-\frac{b(r)}{r}\right)^{-1}$, where $a(r)$ and $b(r)$ respectively denote the redshift and shape functions.
	From Eq. (\ref{Eq.8}), the Finsler metric and its reciprocal can be written as
	\begin{equation}\label{Eq.9}
		g_{\mu\nu}=diag\bigg(e^{v(r)}, -e^{\lambda(r)}, -r^2\bar{g}_{ij}\bigg),
	\end{equation}
	\begin{equation}\label{Eq.10}
		g^{\mu\nu}=diag\bigg(e^{-v(r)}, -e^{-\lambda(r)}, -r^{-2}\bar{g}^{ij}\bigg),
	\end{equation}
	where $ \bar{g}_{ij}$ and $ \bar{g}^{ij}$ are derived from $\bar{F}$ and the indices $i$, $j$ are assigned to the angular coordinates $\theta$, $\phi$. Since $\bar{F}$ represents the two-dimensional Finsler space. So we consider $\bar{F}^2$ as follows
	\begin{equation}\label{Eq.11}
		\bar{F}^2=y^\theta y^\theta + \chi(\theta, \phi)y^\phi y^\phi.
	\end{equation}
	We may obtain the Finsler metric for the two-dimensional Finsler structure  $\bar{F}$ as
	\begin{equation}\label{Eq.12}
		\bar{g}_{ij}=diag\bigg(1, \chi (\theta, \phi)\bigg),
	\end{equation}
	\begin{equation}\label{Eq.13}
		\bar{g}^{ij}=diag\bigg(1, \frac{1}{\chi (\theta, \phi)}\bigg).
	\end{equation}
	We have now determined the geodesic spray coefficients for $\bar{F}^2$ using Eq. (\ref{Eq.4}) as
	\begin{equation*}
		\bar{G}^\theta=-\frac{1}{4}\frac{\partial\chi}{\partial\theta}y^\phi y^\phi,
	\end{equation*}
	\begin{equation*}
		\bar{G}^\phi=-\frac{1}{4\chi}\left(2\frac{\partial\chi}{\partial\theta}y^\phi y^\theta + \frac{\partial\chi}{\partial\phi}y^\phi y^\phi\right).
	\end{equation*}
	Referring to the formula provided in the source mentioned as \cite{Paulb}, we can determine the $\bar{R}ic$ of the Finslerian structure denoted as $\bar{F}$ in the following manner
	\begin{equation}\label{Eq.14}
		\bar{R}ic=\frac{1}{2\chi}\left[-\frac{\partial^2\chi}{\partial\theta^2}+\frac{1}{2\chi}\left(\frac{\partial \chi}{\partial \theta}\right)^2\right].
	\end{equation}
	Given that $\bar{R}ic=\gamma$ represents flag curvature, we can address the differential equation Eq. (\ref{Eq.14}) within the context of the Finsler structure $\bar{F}$ under three distinct scenarios when $(\gamma>0, \gamma=0, \gamma<0)$. In our investigation, we maintain $\gamma$ as a constant value, and subsequently, we derive solutions for the differential equation in each of these situations.
	\begin{equation}\label{Eq.15}
		\bar{F}^2=y^\theta y^\theta +C \sin^2(\sqrt{\gamma}\theta)y^\phi y^\phi ~~~~~~~~  \textrm{for} ~~ (\gamma>0),
	\end{equation}
	\begin{equation}\label{Eq.16}
		\bar{F}^2=y^\theta y^\theta +C \theta^2 y^\phi y^\phi   ~~~~~~~~~~~~~~~~~~~~~~ \textrm{for} ~~ (\gamma=0),
	\end{equation}
	\begin{equation}\label{Eq.17}
		\bar{F}^2=y^\theta y^\theta +C \sinh^2(\sqrt{-\gamma}\theta)y^\phi y^\phi ~~~~ \textrm{for} ~~ (\gamma<0).
	\end{equation}
	Utilizing the solution for $\bar{F}$ under the condition $\gamma>0$, we can express the Finsler structure $F$ as defined in Eq. (\ref{Eq.8}) with the following representation by considering $C=1$= constant
	\begin{equation}\label{Eq.18}
		F^2=e^{v(r)}y^t y^t-e^{\lambda(r)} y^r y^r-r^2 y^\theta y^\theta-r^2\sin^2\theta y^\phi y^\phi + r^2\sin^2\theta y^\phi y^\phi-r^2\sin^2(\sqrt{\gamma}\theta)y^\phi y^\phi.
	\end{equation}
	Now the Finsler structure $F$ in Eq. (\ref{Eq.18}) can be rewritten in the form of
	\begin{equation}\label{Eq.19}
		F^2=\alpha^2+r^2f(\theta)y^\phi y^\phi,
	\end{equation}
	where $\alpha$ represents Riemannian metric and $f(\theta)=\sin^2\theta-\sin^2(\sqrt{\gamma}\theta)$. Hence
	\begin{equation}\label{Eq.20}
		F=\alpha \sqrt{1+\frac{r^2f(\theta)y^\phi y^\phi}{\alpha^2}},
	\end{equation}
	suppose $b_\phi=r\sqrt{f(\theta)}$, we get
	\begin{equation}\label{Eq.21}
		F=\alpha\phi(s), ~~~~~\phi(s)=\sqrt{1+s^2},
	\end{equation}
	where $s=\frac{b_\phi y^\phi}{\alpha}=\frac{\beta}{\alpha}, \beta$ is one form, as an outcome. We can conclude that the $F$ is one of the $(\alpha, \beta)$ Finsler metric.\\
	
	In the Finsler space to determine the Killing equation $K_V(F)=0$ as described in \cite{Li}, we examine the transformation of the Finsler structure in the following manner,
	\begin{equation}\label{Eq.22}
		\left(\phi(s)-s\frac{\partial \phi(s)}{\partial s}\right)K_V(\alpha)+\frac{\partial \phi(s)}{\partial s} K_V(\beta)=0,
	\end{equation}	
	where
	\begin{align*}
		K_V(\alpha)&=\frac{1}{2\alpha}(V_{\mu|\nu}+V_{\nu|\mu})y^\mu y^\nu,\\
		K_V(\beta)&=\left(V^\mu \frac{\partial b_\nu}{\partial x^\mu}+ b_\mu\frac{\partial V^\mu}{\partial x^\nu}\right) y^\nu.
	\end{align*}
	
	" $\mid$ " stands for the covariant derivative with regard to the Riemannian metric $\alpha$.
	
	\begin{equation*}
		\alpha K_V(\alpha)+\beta K_V(\beta)=0,
	\end{equation*}
	as a result
	\begin{equation}\label{Eq.23}
		K_V(\alpha)=0 ~~ \textrm{and} ~~ K_V(\beta)=0,
	\end{equation}
	or
	\begin{equation}\label{Eq.24}
		V_{\mu|\nu}+V_{\nu|\mu}=0, ~~~~~\textrm{and} ~~~~V^\mu\frac{\partial b_\nu}{\partial x^\mu}+b_\mu \frac{\partial V^\mu}{\partial x^\nu}=0.
	\end{equation}
	The presence of the Killing equation leads to the breaking of isometric symmetry in Riemannian space-time.\\
	Now, we obtain the Finsler metric as
	\begin{equation}\label{Eq.25}
		g_{\mu\nu}=diag\bigg(e^{v(r)}, -e^{\lambda(r)}, -r^2, -r^2\sin^2\sqrt{\gamma}\theta\bigg),
	\end{equation}
	\begin{equation}\label{Eq.26}
		g^{\mu\nu}=diag\left(e^{-v(r)}, -e^{-\lambda(r)}, -\frac{1}{r^2}, \frac{1}{-r^2\sin^2\sqrt{\gamma}\theta}\right).
	\end{equation}
	In this context, it is serious to note that the positive flag curvature $(\gamma>0)$ has a significant effect on the field equations
	that emerge within Finslerian space-time.
	
	Another way to calculate the geodesic spray coefficient for the Finsler structure represented as $F$ is through Eq. (\ref{Eq.8}), yielding the following expression,
	\begin{align*}
		G^t&=\frac{v'(r)}{2}y^t y^r,\\
		G^r&=\frac{\lambda'(r)}{4}y^r y^r + \frac{e^{v(r)}v'(r)}{4e^{\lambda(r)}}y^t y^t-\frac{r}{2e^{\lambda(r)}} \bar{F}^2, \\
		G^\theta &=\frac{1}{r}y^\theta y^r+\bar{G}^\theta,\\
		G^\phi &=\frac{1}{r}y^\phi y^r+\bar{G}^\phi,
	\end{align*}
	where prime $(')$ signifies the derivative with respect to r.
	
	Einstein introduced the concept of the geometry-matter relationship, which describes gravity as space-time curvature. The Finslerian gravitational field equation is given by
	\begin{equation}\label{Eq.29}
		G_{\mu\nu}=8\pi_F G T_{\mu\nu}.
	\end{equation}
	
	In the realm of Finsler space-time, the formulation of the Einstein field equation takes on the following expression
	\begin{equation}\label{Eq.27}
		G_{\mu\nu}=Ric_{\mu\nu}-\frac{1}{2}g_{\mu\nu}S,
	\end{equation}
	\begin{equation}\label{Eq.28}
		S=g^{\mu\nu}Ric_{\mu\nu},
	\end{equation}
	in which $ G_{\mu\nu}$, $S$, and $Ric_{\mu\nu}$ represent the Einstein tensors, scalar curvature, and Ricci tensor, respectively, and $T_{\mu\nu}$ stands for energy-momentum tensor, $4\pi_F$ is the volume of the Finsler structure $\bar{F}$  in two dimensions and $G=1$ \cite{Li1}.\\
	Another approach to calculating the Einstein tensors in Finsler space-time is as follows,
	\begin{align}\label{Eq.30}
		G^t_t&=\frac{\lambda'}{r e^\lambda}-\frac{1}{r^2e^\lambda}+\frac{\gamma}{r^2}, \nonumber\\
		G^r_r&=-\frac{v'}{re^\lambda e^v}-\frac{1}{r^2e^\lambda}+\frac{\gamma}{r^2}, \nonumber \\
		G^\theta_\theta &=G^\phi_\phi=-\frac{v''+{v'}^2}{2e^\lambda}-\frac{v'}{2re^\lambda}+\frac{\lambda'}{2re^\lambda}+\frac{v'}{4e^\lambda}(\lambda'+v').
	\end{align}
	Let's consider the generalized anisotropic energy-momentum tensor as \cite{Mak}
	\begin{equation}\label{Eq.31}
		T^\mu_\nu=(\rho+p_t)u^\mu u_\nu+(p_r-p_t)\eta^\mu \eta_\nu-p_t \delta^\mu_\nu,
	\end{equation}
	In this context, $u^\mu$ represents the four-velocity while $\eta^\mu$ signifies a space-like unit vector. It is worth emphasizing that $u^\mu u_\mu =1$ and $\eta^\mu \eta_\mu=-1$. Furthermore, $\rho$, $p_r$, and $p_t$ correspond to the energy density, radial pressure, and transverse pressure respectively.\\
	
	We can calculate the gravitational field equations from Eq. (\ref{Eq.29}) in Finsler space-time using components of Einstein tensors  and energy-momentum tensors as shown below,
	\begin{equation}\label{Eq.32}
		8\pi_F\rho=\frac{\lambda'}{r e^\lambda}-\frac{1}{r^2e^\lambda}+\frac{\gamma}{r^2},
	\end{equation}
	\begin{equation}\label{Eq.33}
		-8\pi_Fp_r=-\frac{v'}{re^\lambda }-\frac{1}{r^2e^\lambda}+\frac{\gamma}{r^2},
	\end{equation}
	\begin{equation}\label{Eq.34}
		-8\pi_Fp_t=-\frac{v''+{v'}^2}{2e^\lambda}-\frac{v'}{2re^\lambda}+\frac{\lambda'}{2re^\lambda}+\frac{v'}{4e^\lambda}(\lambda'+v').
	\end{equation}

\subsection{The shape function}\label{subsec2.1}
The shape function is a crucial element in the traversable wormhole analysis as it characterizes the geometric properties of space-time curvature in the vicinity of these theoretical structures. Let's delve deeper into the shape functions derived for both models under conformal motion, which is represented as an arbitrary function of the radial coordinate $r$ and adheres to the following criteria \cite{Morris}
\begin{itemize}
	\item  The radial coordinate $r$ in this context exhibits a monotonically increasing function with $b(r)>0$ $\forall$ $r>r_0$.
	\item To fulfill the throat condition, the shape function $b(r)$ requires a specific point where it remains unchanged $b(r_0)=r_0$, and $b(r_0)-r_0<0$ at thr throat radius $r_0$.
	\item   To ensure the validity of wormhole solutions, a key requirement is that the expression $b'(r_0)<1$ at $r=r_0$. This condition is known as the flaring-out condition and ensures that the wormhole throat is expanding.
	\item In regions where $r > r_0$,  $\frac{b(r)}{r}$ should approach zero and the function $a(r) \rightarrow 0$  as $|r|\rightarrow \infty$. It is also imperative to ensure that $\frac{b(r)}{r}<1$.
\end{itemize}
To enable the traversability of the wormhole, it is vital to guarantee the absence of horizons along the pathway. It requires satisfying crucial criteria, which include the flaring-out condition at the throat and the asymptotic flatness criteria.

\subsection{The proper radial distance}\label{subsec2.2}
When exploring the concept of wormholes in space-time to teach GR, it is highlighted that while the radial measurement r behaves irregularly near the throat, the proper radial distance $l(r)$ needs to remain stable throughout space-time. This means that $l(r)$ must stay finite everywhere, indicating that the expression $1-b(r)/r$ must consistently be positive or zero across space-time. This is crucial because a value of zero would make the proper radial distance infinite, although it doesn't necessarily imply an infinite change in distance \cite{Morris}. The radial coordinate $r$ often encounters issues or peculiar behavior near the throat of the wormhole, such as coordinate singularities or unusual behaviors in certain coordinate systems. The proper radial distance, on the other hand, takes into account the physical effects and provides a more meaningful measure of distance within the Finsler wormhole geometry, and it can be written as
\begin{equation}\label{Eq.55a}
	l(r)=\pm \int_{r_0}^{r}\frac{dr}{\sqrt{1-\frac{b(r)}{r}}}.
\end{equation}

\subsection{The radial velocity}
Understanding the radial velocity of a wormhole holds paramount significance in comprehending its traversability and temporal dynamics within the framework of space-time. The radial velocity delineates the rate of change in the wormhole's radial coordinate, 'r,' particularly near its throat, where space-time curvature is pronounced. This velocity factor critically influences the stability and geometry of the wormhole, impacting the feasibility of traversing it and the potential for sustaining a passage between disparate regions of space-time.	Analyzing the radial velocity offers insights into the energy conditions required to uphold a traversable wormhole, shedding light on the fundamental constraints and physical possibilities governing these hypothetical structures. The radial velocity of a traveler moving through a wormhole refers to the component of their velocity that is directed along the radial direction of the Finslerian wormhole's space-time, and it can be expressed as follows,
\begin{equation}\label{Eq.57a}
	V(r)= \pm \frac{dr}{e^\Phi dt\sqrt{1-\frac{b(r)}{r}}}.
\end{equation}

	
	\section{Formulation of a Finslerian wormhole incorporating conformal motion}\label{sec.3}
	In Finsler space, the metric tensor field $g$ is associated with the definition of  CKVs denoted by $\xi$. These CKVs are determined through the Lie infinitesimal operator $\mathcal{L}_{\hat{\xi}}$, resulting in the following representation for $\xi$ in the Finsler space-time \cite{Joharinad}:
	\begin{equation}\label{Eq.1}
		\mathcal{L}_{\hat{\xi}}g_{\mu\nu}=\nabla_\mu\xi_\nu+\nabla_\nu\xi_\mu+2y^n(\nabla_n\xi^\alpha)\mathcal{C}_{\alpha\mu\nu}=\psi(r)g_{\mu\nu},
	\end{equation}
	where
	\begin{equation}\label{Eq.71}
		\nabla_\mu\xi_\nu=\frac{\partial\xi_\nu}{\partial x^\mu}-G^\alpha_\mu\frac{\partial\xi_\nu}{\partial y^\alpha}-\Gamma^\alpha_{\mu\nu}\xi_\alpha ,
	\end{equation}
	and
	\begin{equation}\label{Eq.72}
		\mathcal{C}_{\alpha\mu\nu}=\frac{F}{4}\frac{\partial^3F^2}{\partial y^\alpha \partial y^\mu \partial y^\nu},
	\end{equation}
	in this expression $\mathcal{L}$ represents the Lie derivative operator and $\psi$ denotes the conformal factor and CKVs as $\hat{\xi}=\xi^\mu(r)\frac{\partial}{\partial x^\mu}+y^\nu(\partial_{\nu}\xi^{\mu})\frac{\partial}{\partial y^\mu}$.  $\mathcal{C}_{\alpha\mu\nu}$ is a Cartan connection in the context of the Finslerian wormhole structure Eq. (\ref{Eq.8}), the observation that $\mathcal{C}_{\alpha\mu\nu}$ equals zero for all indices $\alpha, \mu, \nu$ implies that this condition characterizes the geometric properties of the Finslerian wormhole space-time. Consequently, we can deduce that the Cartan connection becomes null within the Finslerian wormhole space-time. The CKV $\xi$ yields the conformal symmetry. An important insight is that neither $\xi$ nor $\psi$ are required to be static when dealing with a static metric. The outcomes of Eq. (\ref{Eq.1}) vary depending on the value of $\psi$. For instance, when $\psi=0$, it results in the Killing vector indicating asymptotic flatness of space-time. When $\psi$ is a constant, it yields a homothetic vector, and when $\psi=\psi(x, t)$, it gives rise to conformal vectors \cite{Herrera}.	With $\xi^{\mu}=g^{\mu\nu}\xi_\nu$, Eq. (\ref{Eq.1}) produced the following expression \cite{Bohmer}
	\begin{align*}
		\xi^r v'&=\psi,\\
		\xi^t&=A_1,\\
		\xi^r&=\frac{\psi r}{2},\\
		\psi^r\lambda'+2\psi^r_{,r}&=\psi.
	\end{align*}
	Here, in the context of our current study, the symbol $r$ represents spatial coordinates while $t$ represents temporal coordinates. In our investigation, the conformal factor $\psi$ is treated as a function of the radial coordinate $r$.
	\begin{equation}\label{Eq.36}
		e^v=e^{2a(r)}=A_2^2r^2,
	\end{equation}
	\begin{equation}\label{Eq.37}
		e^\lambda=\left(1-\frac{b(r)}{r}\right)^{-1}=\left(\frac{A_3}{\psi}\right)^2,
	\end{equation}
	\begin{equation}\label{Eq.38}
		\xi^i=A_1\delta^i_t+\left(\frac{\psi r}{2}\right)\delta^i_r,
	\end{equation}
	where $A_1$, $A_2$, and $A_3$ are the integration constants, using Eqs. (\ref{Eq.36}-\ref{Eq.38}) rewriting the field equations (\ref{Eq.32}-\ref{Eq.34}) as
	\begin{equation}\label{Eq.39}
		8\pi_F\rho=\frac{1}{r^2}\left[\gamma-\frac{\psi^2}{A^2_3}\right]-\frac{2\psi'\psi}{rA_3^2},
	\end{equation}
	\begin{equation}\label{Eq.40}
		-8\pi_F p_r=\frac{1}{r^2}\left[\gamma-\frac{3\psi^2}{A^2_3}\right],
	\end{equation}
	\begin{equation}\label{Eq.41}
		8\pi_F p_t=\frac{\psi^2}{r^2A^2_3}+\frac{2\psi \psi'}{r A^2_3}.
	\end{equation}
\section{Wormholes solution to Finslerian gravity}\label{sec.4}
In this context, the objective is to discover a solution for a Finslerian wormhole by employing the field equations of Finsler space-time. To do this, we introduce a particular EoS that establishes a linear connection between the radial pressure ($p_r$) and the energy density ($\rho$). This EoS is characterized by the parameter $\omega$, which governs the relationship between $p_r$ and $\rho$. Using the linear EoS, it can be expressed as follows
	\begin{equation}\label{Eq.42}
		p_r=\omega\rho.
	\end{equation}
	The EoS provides insights into the characteristics of the fluid responsible for the ongoing expansion of the universe driven by acceleration. This EoS provides a way to characterize different types of cosmic fluids based on their behavior under varying conditions. In wormhole studies involving baryonic and non-baryonic fluids, the equation of state parameter $(\omega)$ plays a significant role in characterizing the behavior of these fluids. For a baryonic fluid, $ 0 \leq \omega < 1 $. This range signifies the behavior of normal matter that obeys classical physics, particularly:
	\begin{itemize}
		\item[*] For dust, $\omega=0$.
		\item[*] For radiation or non-relativistic matter such as cold dark matter (CDM),$\omega=\frac{1}{3}$.
		\item[*] For stiff fluid, $\omega=1$.
	\end{itemize}
	On the other hand, non-baryonic fluids often involve $\omega$ values that differ from those in the baryonic case.
	
	\begin{itemize}
		\item[*] For dark energy $-1\leq\omega\leq -\frac{1}{3}$, if it behaves like a cosmological constant (as in the $\Lambda$CDM model), $\omega = -1$.
		\item[*] For Phantom fluid $\omega<-1$.
	\end{itemize}
	The specific value of $\omega$ influences the fluid's properties, like pressure and density, affecting how it interacts within the context of a wormhole. Upon solving Eq. (\ref{Eq.42}) utilizing Eqs. (\ref{Eq.39}) and (\ref{Eq.40}) we arrive at the following result
	\begin{equation}\label{Eq.43}
		\psi^2=\frac{K_1(\omega+3)r^{-\frac{\omega+3}{\omega}}+(\omega+1)\gamma A_3^2}{\omega+3},
	\end{equation}
	where $K_1$ denotes the integration constant, further the shape function $b(r)$ can be calculated from Eq. (\ref{Eq.37}) by using Eq. (\ref{Eq.43}) resulting in the following expression
	\begin{equation}\label{Eq.44}
		b(r)=-r\left[\frac{K_1(\omega+3)r^{-\frac{\omega+3}{\omega}}+((\gamma-1)\omega+\gamma-3)A_3^2}{(\omega+3)A_3^2}\right].
	\end{equation}
	Utilizing Equation (\ref{Eq.43}) we can express the field equations (\ref{Eq.39}-\ref{Eq.41}) in the following manner
	\begin{equation}\label{Eq.45}
		\rho=\frac{3K_1(\omega+3)r^{-\frac{\omega+3}{\omega}}+2\gamma A_3^2\omega}{(\omega+3)r^2A_3^2\omega8\pi},
	\end{equation}
	\begin{equation}\label{Eq.46}
		p_r=\frac{3K_1(\omega+3)r^{-\frac{\omega+3}{\omega}}+2\gamma A_3^2\omega}{(\omega+3)r^2A_3^2 8\pi},
	\end{equation}
	\begin{equation}\label{Eq.47}
		p_t=\frac{-3K_1(\omega+3)r^{-\frac{\omega+3}{\omega}}+\gamma A_3^2\omega(\omega+1)}{(\omega+3)r^2A_3^2 \omega 8\pi}.
	\end{equation}
	Furthermore, we can represent the other components by employing Equations (\ref{Eq.45}-\ref{Eq.47}) as shown below,
	\begin{equation}\label{Eq.48}
		\rho+p_r=\frac{3(\omega+1)\left[K_1(\omega+3)r^{-\frac{\omega+3}{\omega}}+\frac{2}{3}\gamma A^2_3\omega\right]}{(\omega+3)r^2A_3^2 \omega 8\pi},
	\end{equation}
	\begin{equation}\label{Eq.49}
		\rho-p_r=-\frac{3(\omega-1)\left[K_1(\omega+3)r^{-\frac{\omega+3}{\omega}}+\frac{2}{3}\gamma A^2_3\omega\right]}{(\omega+3)r^2A_3^2 \omega 8\pi},
	\end{equation}
	\begin{equation}\label{Eq.50}
		\rho+p_t=\frac{\gamma}{8\pi r^2},
	\end{equation}
	\begin{equation}\label{Eq.51}
		\rho-p_t=\frac{6K_1(\omega+3)r^{-\frac{\omega+3}{\omega}}-\gamma A^2_3\omega(\omega-1)}{(\omega+3)r^2A_3^2 \omega 8\pi},
	\end{equation}
	\begin{equation}\label{Eq.52}
		p_t-p_r=\frac{-3K_1(\omega+3)(\omega+1)r^{-\frac{\omega+3}{\omega}}-\gamma A^2_3\omega(\omega-1)}{(\omega+3)r^2A_3^2 \omega 8\pi},
	\end{equation}
	\begin{equation}\label{Eq.53}
		\rho+p_r+2p_t=\frac{3K_1(\omega+3)(\omega-1)r^{-\frac{\omega+3}{\omega}}+4\gamma A^2_3\omega(\omega+1)}{(\omega+3)r^2A_3^2 \omega 8\pi}.
	\end{equation}	
	From eq. (\ref{Eq.57a}), we obtain the radial velocity
	\begin{equation}\label{Eq.57aa}
		v  = \frac{\sqrt{\omega+3}A_3}{\sqrt{K_1(\omega+3)r^{\frac{-\omega-3}{\omega}}A_3^2\gamma(\omega+1)}}.
	\end{equation}

\subsection{Physical  viability of the Wormhole solution with baryonic fluid}\label{subsec4.1}
This section delved into the physical behavior of the wormhole, highlighting its importance for its
credibility and stability. We study the physical affirmation of the proposed solutions for the wormhole
via shape function, proper radial distance, radial velocity, energy density, pressure, energy conditions, and anisotropy with baryonic fluids models (dust, stiff, and radiation fluid) that are feasible and well-behaved in gravitational theory. Furthermore, the graphical illustrations are also discussed in detail.

\begin{figure}[hptb]
	\begin{center}
	        \includegraphics[scale=0.4]{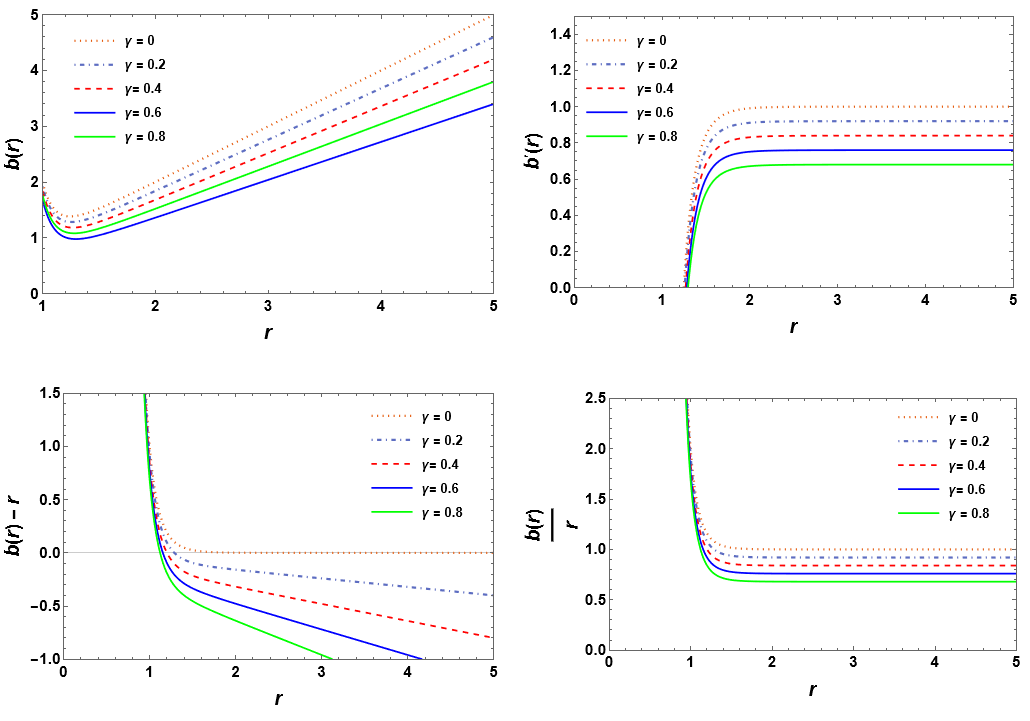}
  			\caption{\label{fig1bb}The graphs illustrate the characteristics of the shape function described by Eq. (\ref{Eq.44}) across various  values of $\gamma>0$ for baryonic fluids, specifically at $\omega = 1/3$ (representing radiation or non-relativistic matter) for $b(r),b'(r), b(r) - r , $ and $ b(r)/r$, while maintaining with $K_1=-1$,  $A_3=1$.}
	\end{center}
\end{figure}

\begin{figure}[hptb]
	\begin{center}
		\includegraphics[scale=0.4]{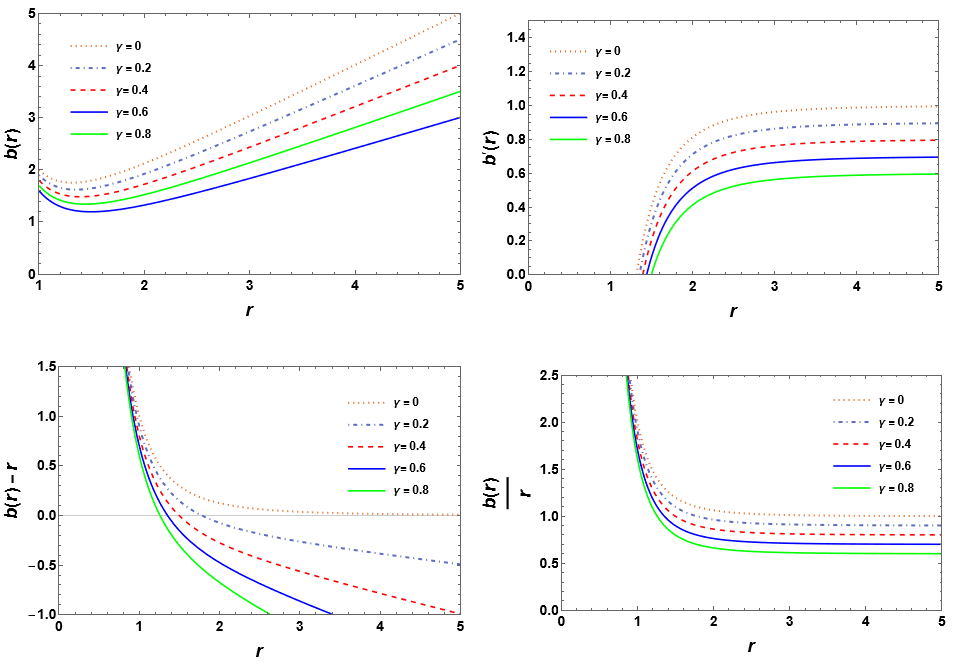}
		\caption{\label{fig1}The graphs illustrate the characteristics of the shape function described by Eq. (\ref{Eq.44}) across various  values of $\gamma>0$ for baryonic fluids, specifically at $\omega = 1$ (representing a stiff fluid) for $b(r), b'(r), b(r)-r$, and $ b(r)/r$, while maintaining with $K_1=-1$,  $A_3=1$.}
	\end{center}
\end{figure}

	In our current investigation, we have computed the derived shape function pertaining to baryonic fluid, specifically for radiation at $\omega=1/3$ in Fig. (\ref{fig1bb}) and for a stiff fluid at $\omega=1$ in Fig. (\ref{fig1}). These plots, varying with the Finsler parameter ($\gamma>0$) under conformal motion, meet all the essential geometric criteria required for a wormhole structure. The value of $b(r)-r$ represents the position of the wormhole throat when it intersects the $r-$axis at the point  $r = r_0$, as depicted in Fig. (\ref{fig1bb}) and (\ref{fig1}). Table (\ref{tbl1}) showcases the throat radius of the Finslerian wormholes for different $\gamma$ values, taking into account both $\omega=\frac{1}{3}$, and $\omega=1$.
	
	\begin{table}[htbp]
		\centering
			\caption{The throat position $r_0$ in a Finslerian wormhole, for radiation $\left(\omega = \frac{1}{3}\right)$ and a stiff fluid  $(\omega=1)$, varies with different values of $\gamma$.}\label{tbl1}
		 \begin{tabular}{lcc}
				\hline
			\bf{${\bf\gamma}$} & \bf{$r_0\left(\omega=\frac{1}{3}\right)$} & \bf{$r_0(\omega=1)$}  \\
				\hline
			0  & 0       & 0 \\	
		    0.2& 1.28734 & 1.77828\\	
			0.4& 1.20112 & 1.49535\\	
			0.6& 1.15339 & 1.35120\\
			0.8& 1.12069 & 1.25743\\
			\hline	
	\end{tabular}
	
	\end{table}

\begin{figure}[hptb]
		\begin{center}
			\mbox{{\includegraphics[scale=0.42]{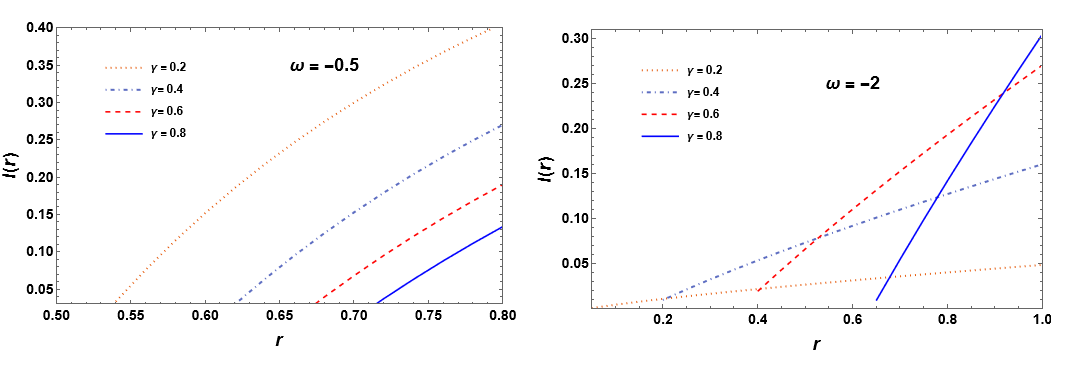}}}
			\caption{\label{fig1aa} The graph displaying the proper radial distance $l(r)$ against $r$ for different $\gamma$ values in baryonic fluid mirrors the description provided in Fig. (\ref{fig1})}
		\end{center}
	\end{figure}

	The integral Eq. (\ref{Eq.55a}) poses a challenge for analytical solutions, so we resort to a numerical method. We obtain values by assigning the lower limit $(r_0)$ as the throat radius from Table (\ref{tbl1}) and the upper limit $(r)$ as a point outside the wormhole throat. This process generates numerical values, allowing us to plot the graph of $l(r)$ versus $r$, shown in Fig. (\ref{fig1aa}). In the wormhole scenario containing baryonic fluid, the proper radial distance consistently increases, showcasing finite values across different $\gamma$ variations.
		
\begin{figure}[hptb]
	\begin{center}
		{\includegraphics[scale=0.4]{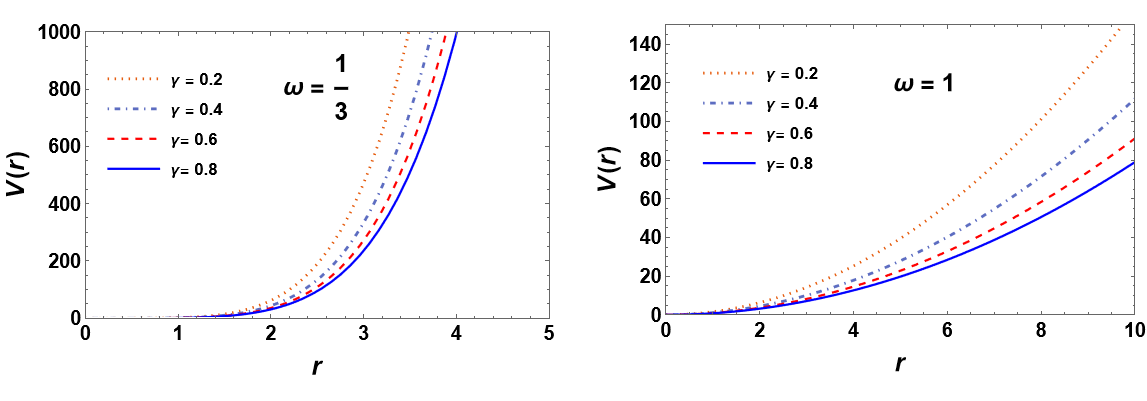}}
		\caption{\label{fig1a} Graph illustrating the radial velocity of baryonic fluid corresponding to radiation $(\omega = 1/3)$ (left). Stiff fluid $(\omega = 1)$ with specified parameters $K_1 = 1,A_3 = 1$ (right).}
	\end{center}
\end{figure}

Using Eq. (\ref{Eq.57aa}), we created a plot depicting the radial velocity $V(r)$ against $r$ for both $\omega=\frac{1}{3}$ and $\omega=1$, as presented in Fig. (\ref{fig1a}). In the context of baryonic fluid, we observed that the radial velocity increases once it crosses the wormhole throat.

Furthermore, we discuss several energy conditions and the approach for forecasting the existence of wormholes. They are particularly significant and useful in explaining traversable wormhole geometry since violations of energy criteria can predict the existence of wormholes. The energy conditions, which include the null energy condition (NEC), dominant energy condition (DEC), weak energy condition (WEC), and strong energy condition (SEC), are defined based on the quantities $\rho$ (energy density), $p_r$ (radial pressure), and $p_t$ (transverse pressure) \cite{Morris}. These conditions can be expressed as follows,
\begin{itemize}
	\item [1.] ${\bf NEC}\Leftrightarrow \rho+p_l\geq 0$, ~~$ \forall ~~l$,
	\item [2.] ${\bf DEC} \Leftrightarrow \rho\geq 0$ and $\rho-p_l\geq0 ~~\forall ~~l$,
	\item [3.] ${\bf WEC} \Leftrightarrow \rho\geq 0$ and $\rho+p_l \geq0 ~~\forall ~~l$,
	\item [4.] ${\bf SEC} \Leftrightarrow \rho+p_l\geq 0$ and $\rho+\sum_l p_l\geq 0 ~~\forall~~ l$.
\end{itemize}
In the current investigation, we delve into the violation of energy conditions, specifically the NEC, DEC, WEC, and SEC. These conditions are fundamental in describing the behavior of energy and different matter in the context of GR and more broadly, in wormhole physics. The presence of violations in these energy conditions is essential in the exotic matter study and its role in the formation and stability of wormholes. Our analysis places particular emphasis on how the parameter $\gamma$ within the Finsler geometry affects the violation of these energy conditions in these two distinct models. To gain a more quantitative perspective, we examined the energy conditions and visualized their patterns graphically.

\begin{figure}[hptb]
	\begin{center}
		{\includegraphics[scale=0.41]{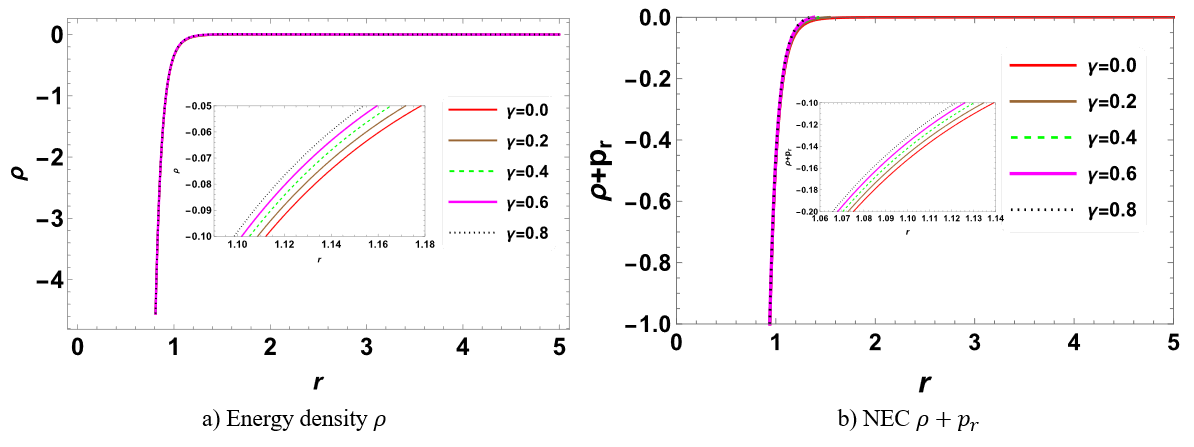}}
		{\includegraphics[scale=0.4]{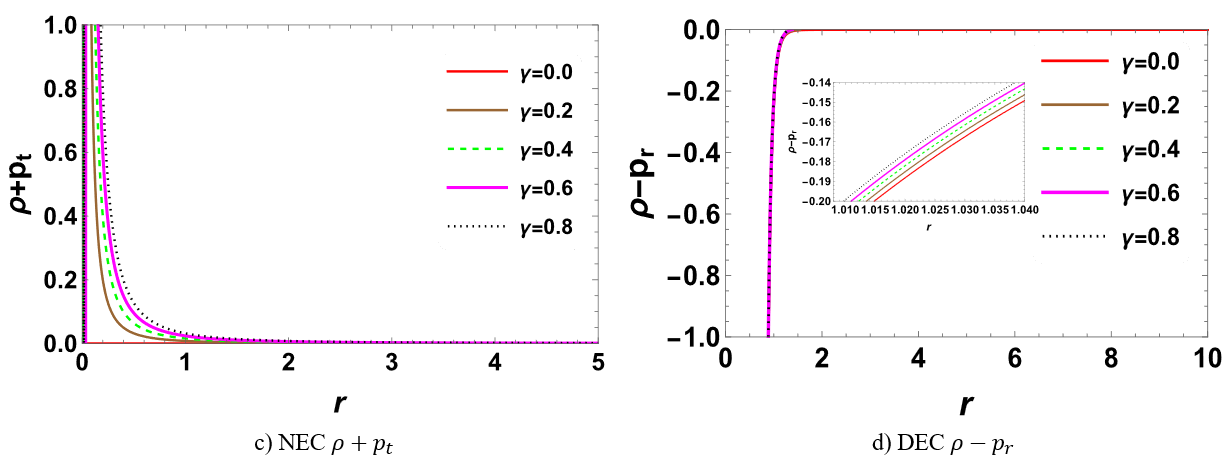}}
		{\includegraphics[scale=0.4]{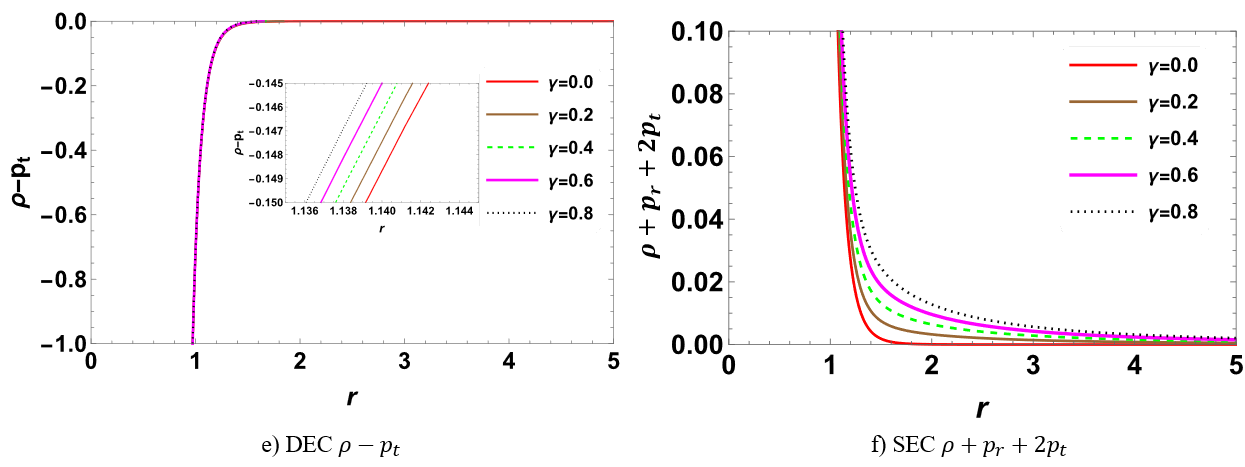}}
		\caption{\label{IMG5} Graphs illustrating the energy conditions, energy density $\rho$, NEC, WEC, DEC, and SEC for baryonic matter, specifically for radiation with parameters $\omega = \frac{1}{3}, \gamma\geq0, K_1 = -1,$ and $A_3 = 1.$}
	\end{center}
\end{figure}

\begin{figure}[hptb]
	\begin{center}
		{\includegraphics[scale=0.41]{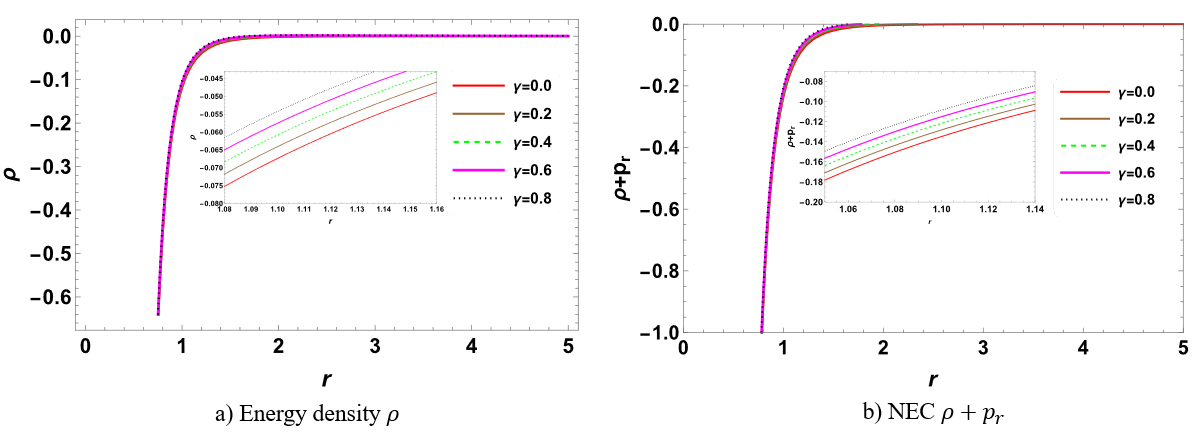}}
			{\includegraphics[scale=0.4]{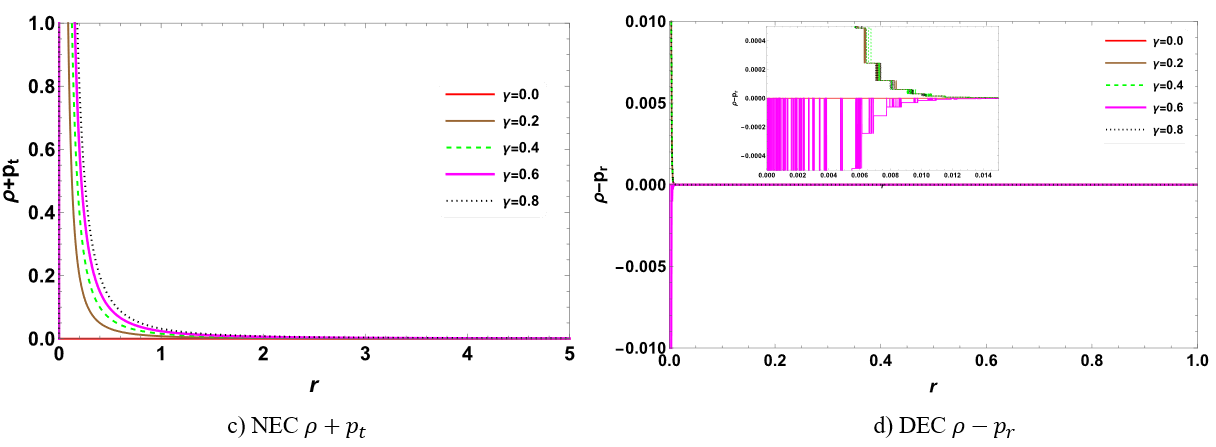}}
				{\includegraphics[scale=0.4]{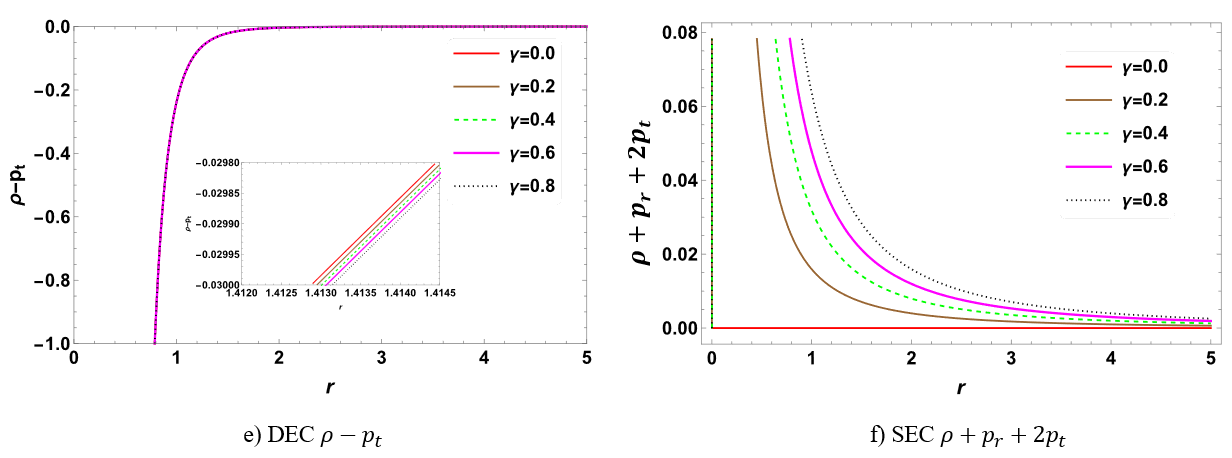}}
		\caption{\label{IMG6}  Graphs illustrating the energy conditions, energy density $\rho$, NEC, DEC, WEC, and SEC for baryonic matter, specifically for stiff fluid with parameters $\omega = 1, \gamma\geq0, K_1=-1,$ and $A_3 = 1.$}
	\end{center}
\end{figure}
In this context, ``dust" isn't literal; it's a cosmological term simplifying matter distribution. Near a wormhole, radiation or non-relativistic matter like cold dark matter is crucial. Their influence can stabilize or destabilize the wormhole, impacting its structure. Understanding this behavior is vital for assessing wormhole durability. Including a "stiff fluid" in models can counterbalance gravity's effects, aiding wormhole formation. This exotic matter's extreme pressure might sustain the wormhole's structure. Radiation or stiff fluid, in theory, can be inferred when energy conditions like WEC and NEC seem violated in specific scenarios.
\begin{itemize}
	\item  The violation of the NEC is a crucial aspect of maintaining a traversable wormhole, as it permits the presence of baryonic matter with specific energy and pressure properties. The NEC imposes a limit on the energy and pressure experienced by an observer moving at the speed of light. In our current analysis, depicted in Figs. (\ref{IMG5}) and (\ref{IMG6}), we observe the violation of the NEC within the wormhole's throat for values of $\gamma\geq0$. Specifically, violations occur in terms of $\rho+p_r\leq0$  [see Fig. \ref{IMG5}b, \ref{IMG6}b], but valid in terms of $\rho+p_t\geq0$ [see Fig. \ref{IMG5}c, \ref{IMG6}c] for both radiation and stiff fluid, respectively results align with \cite{Ksh}.
	
	\item  According to Hawking's description \cite{Hawking}, the WEC asserts that all physical systems should exhibit positive energy densities from the viewpoint of any observer within space-time. In this analysis, we note a violation of the WEC, as indicated by Fig. (\ref{IMG5}) and Fig. (\ref{IMG6}). For radiation and stiff fluid, the violation is observed in terms of $\rho\leq0$ [see Fig. \ref{IMG5}a, \ref{IMG6}a], $\rho+p_r\leq0$ [see Fig. \ref{IMG5}b, \ref{IMG6}b], and valid interms of $\rho+p_t\geq0$ [see Fig.\ref{IMG5}c, \ref{IMG6}c],  within the range $\gamma\geq0$. This discrepancy suggests the presence of either radiation or stiff fluid near the throat of the Finslerian wormhole under conformal motion.

	\item  The violation of the DEC can be observed from Fig. (\ref{IMG5}) and Fig. (\ref{IMG6}) in terms of $\rho\leq0$ [see Fig. \ref{IMG5}a, \ref{IMG6}a],  $\rho-p_r\leq0$ [see Fig. \ref{IMG5}d, \ref{IMG6}d ($\gamma$=0.6)], and $\rho-p_t\geq0$ [see Fig. \ref{IMG5}e, \ref{IMG6}e] when $\gamma\geq 0$ for both radiation and stiff fluid respectively.
	
	\item The fourth energy condition SEC is valid for $\gamma\geq0$  interms of $\rho+p_r+2p_t$ [see Fig. \ref{IMG5}f, \ref{IMG6}f] near the wormhole throat for radiation and stiff fluid, respectively.

\end{itemize}

\subsection{Physical  viability of the Wormhole solution with non-baryonic fluid}\label{subsec4.2}
The non-baryonic fluids viz Dark energy are hypothetical forms of energy that are thought to permeate space and are associated with the observed accelerated expansion of the universe. Phantom energy is a speculative form of dark energy characterized by an equation of state where pressure is more negative than the energy density, leading to increasingly repulsive effects. The importance attributed to the presence of dark energy or phantom energy near the throat of a wormhole is largely theoretical and tied to exotic matter requirements for stabilizing and maintaining a traversable wormhole. The idea is that these exotic forms of energy might counterbalance the extreme gravitational forces and keep the wormhole open. When energy conditions are violated, it could result in the presence of non-baryonic fluid (Dark energy or phantom energy) near the wormhole throat.
	\begin{figure}[hptb]
		\begin{center}
			{\includegraphics[scale=0.37]{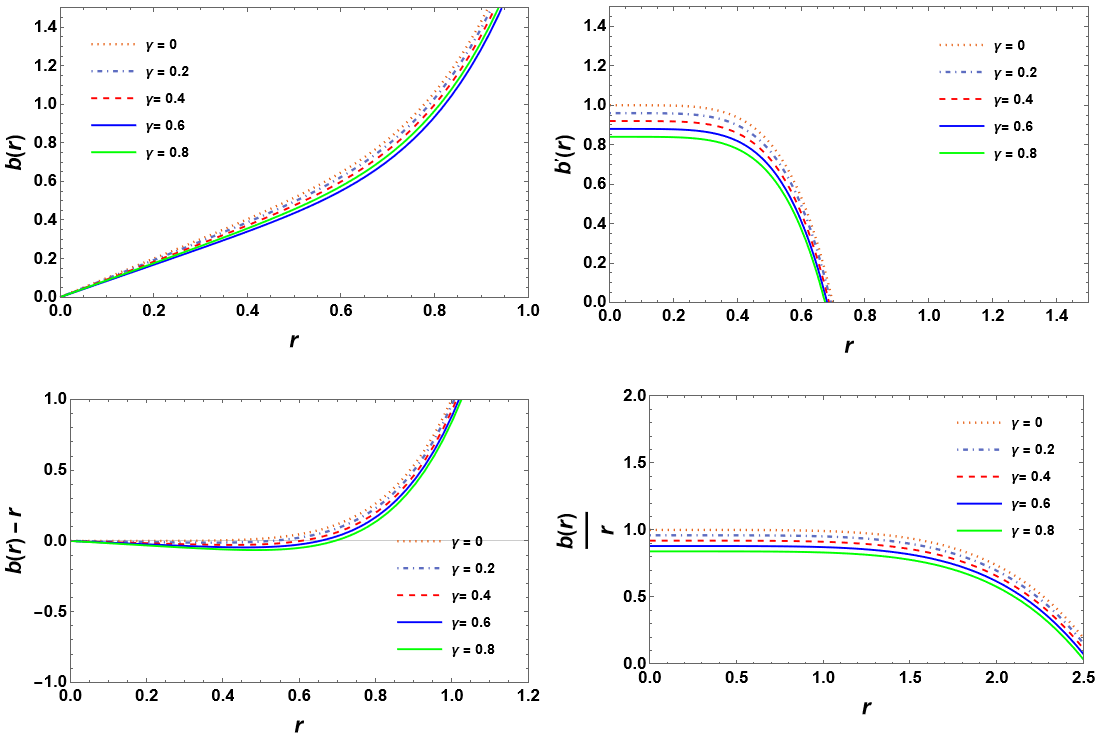}}
					\caption{\label{fig2bb} The graphs illustrate the characteristics of the shape function described by Eq. (\ref{Eq.44}) across various  values of $\gamma>0$ for non-baryonic fluids, specifically at $\omega = -0.5$ (representing dark energy), for $b(r)$, and $b(r)-r$ (with $K_1=-1)$, $b'(r)$, $b(r)/r$ (with $K_1=1)$, maintaining $A_3=1$.}
		\end{center}
	\end{figure}
	\begin{figure}[hptb]
	\begin{center}
		{\includegraphics[scale=0.39]{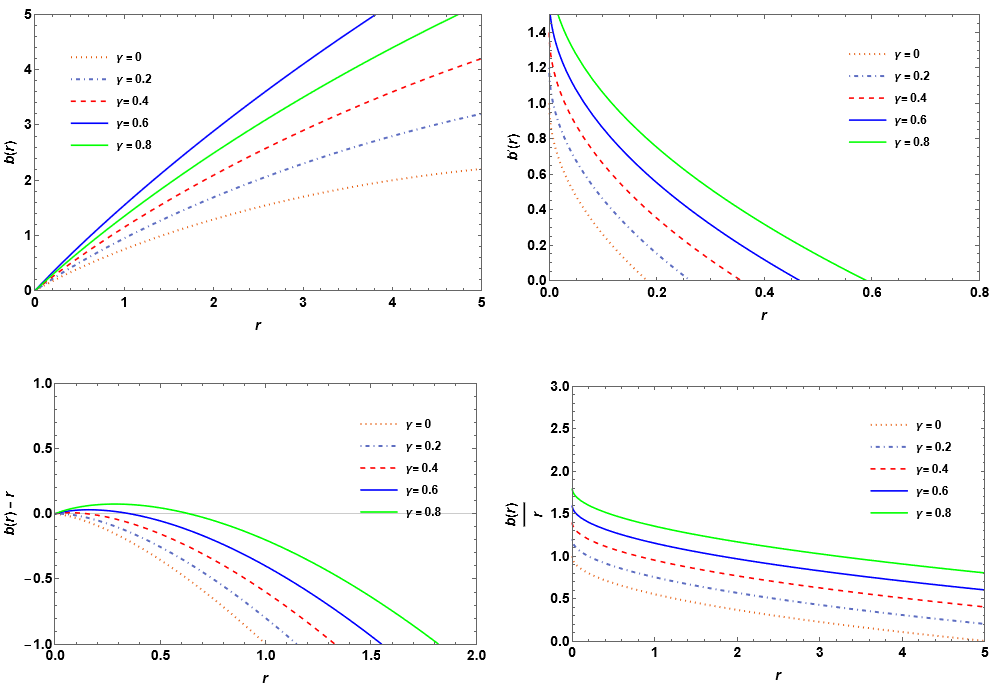}}
		\caption{\label{fig2cc} The graphs illustrate the characteristics of the shape function described by Eq. (\ref{Eq.44}) across various  values of $\gamma>0$ for non-baryonic fluids, specifically at $\omega = -2$ (representing a phantom fluid), for  $b(r)$ ($A_3=2$), and $b'(r)$ $(A_3=0.8)$, $b(r)-r$ $(A_3=1)$, $b(r)/r$ $(A_3=1.5)$, while maintaining $K_1=1$.}
	\end{center}
\end{figure}
We plotted the graphs representing different scenarios of the shape function for non-baryonic matter, particularly highlighting dark energy at $\omega=-0.5$ in Fig. (\ref{fig2bb}) and phantom fluid at $\omega=-2$ in Fig. (\ref{fig2cc}), illustrate that the derived shape function meets all geometric conditions. Table \ref{tbl2} outlines the throat radius of Finslerian wormholes in scenarios of non-baryonic fluid, considering different $\gamma$ values for both $\omega=-0.5$ and $\omega=-2$.

	\begin{table}[htbp]
	\centering
	\caption{The throat position $r_0$ in a Finslerian wormhole, for dark energy $(\omega =-0.5)$ and a phantom fluid  $(\omega=-2)$, for different values of $\gamma$.}\label{tbl2}
	\begin{tabular}{lcc}
		\hline
	 ${\gamma}$ & $r_0 (\omega=-0.5)$ & $r_0(\omega=-2)$  \\
		\hline
		0  &  0      & 0    \\	
		0.2&  0.52530& 0.04 \\	
		0.4&  0.60342& 0.16 \\	
		0.6&  0.65439& 0.36 \\
		0.8&  0.69314& 0.64 \\
		\hline	
	\end{tabular}
	
\end{table}

\begin{figure}[hptb]
	\begin{center}
		\mbox{{\includegraphics[scale=0.39]{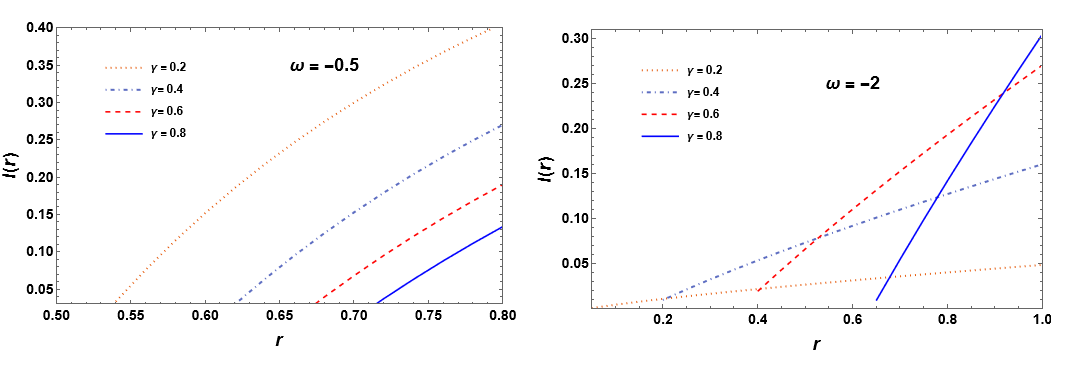}}}
		
		\caption{\label{fig2aa} The graph displaying the proper radial distance $l(r)$ against $r$ for different $\gamma$ values in non-baryonic fluid mirrors the description provided in Fig. (\ref{fig2bb}) and (\ref{fig2cc}).}
	\end{center}
\end{figure}

In Fig. (\ref{fig2aa}), the plot showcases how the radial velocity $l(r)$ versus wormhole radius $r$ for a Finslerian wormhole containing non-baryonic fluid with $\omega=-0.5$ and $\omega=-2$. Across different $\gamma$ values, the plot demonstrates the radial velocity consistently increases with $r$.

\begin{figure}[hptb]
	\begin{center}
		{\includegraphics[scale=0.34]{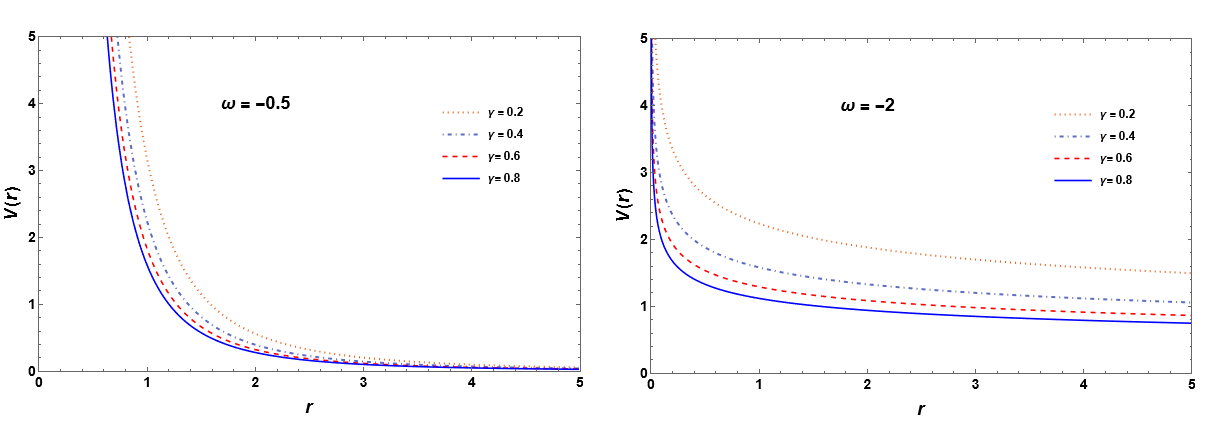}}
			\caption{\label{fig2a} Graph illustrating the radial velocity of non-baryonic fluid corresponding to dark energy $(\omega = -0.5, K_1=1, A_3 = 1)$ (left). Phantom fluid $(\omega = -2, K_1=-1, A_3 = 1)$ (right).}
	\end{center}
\end{figure}

Using the Eq. (\ref{Eq.57aa}), we plotted a graph displaying the radial velocity $V(r)$  against the wormhole radius $r$ for both $\omega=-0.5$ and $\omega=-2$, shown in Fig. (\ref{fig2a}). In the context of non-baryonic fluid, we noticed that the radial velocity $V(r)$ decreases and approaches zero as it crosses the wormhole throat.

\begin{figure}[hptb]
	\begin{center}
		{\includegraphics[scale=0.4]{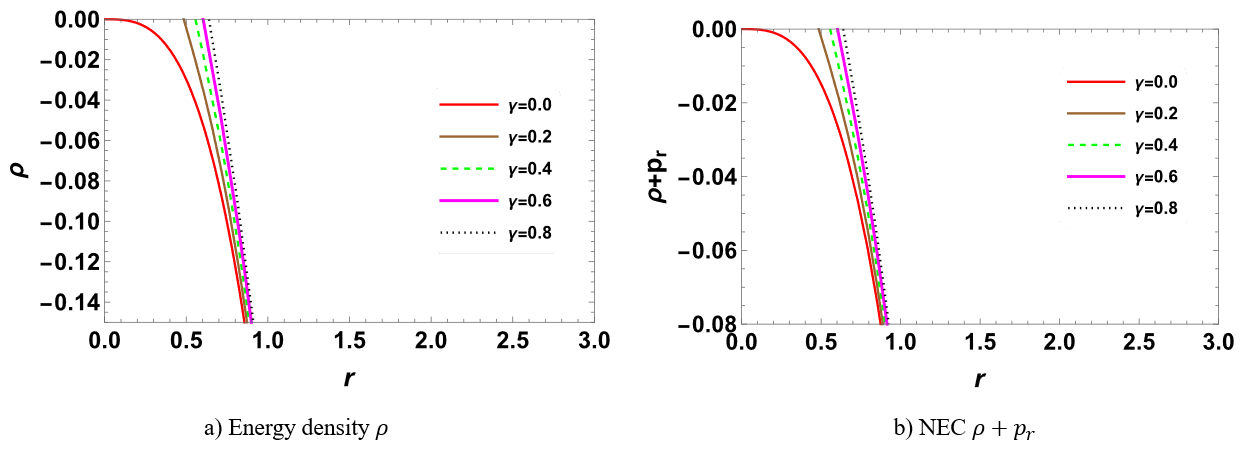}}
		{\includegraphics[scale=0.42]{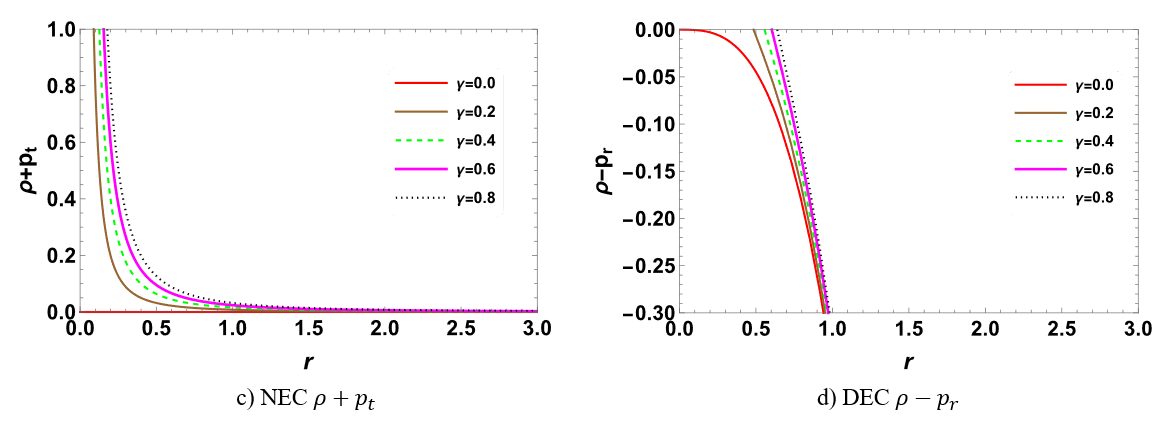}}
		{\includegraphics[scale=0.41]{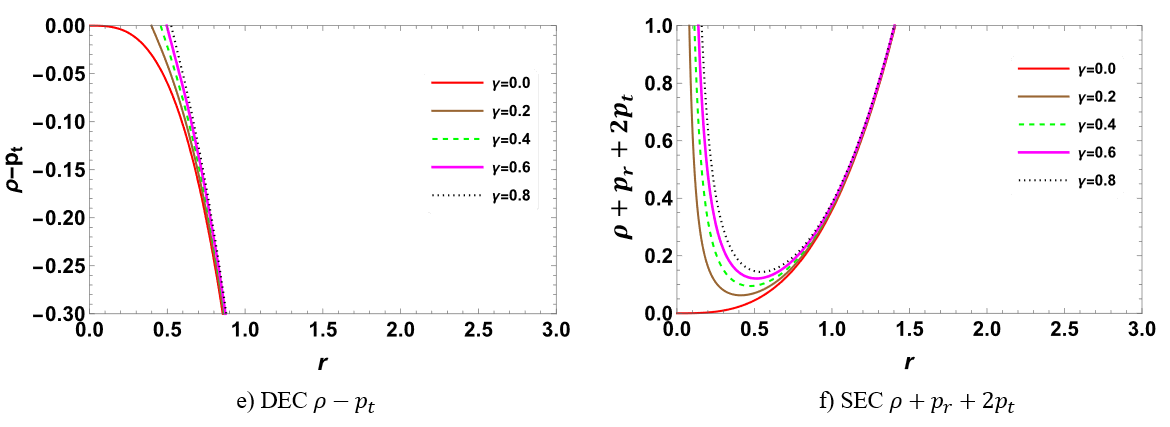}}
		\caption{\label{IMG11} Graphs illustrating the energy conditions radial density $\rho$, NEC, DEC, WEC, and SEC for non-baryonic matter, specifically for dark energy with parameters $\omega = -0.5, \gamma\geq0, K_1=1,$ and $A_3 = 1.$}
	\end{center}
\end{figure}

\begin{figure}[hptb]
		\begin{center}
			{\includegraphics[scale=0.44]{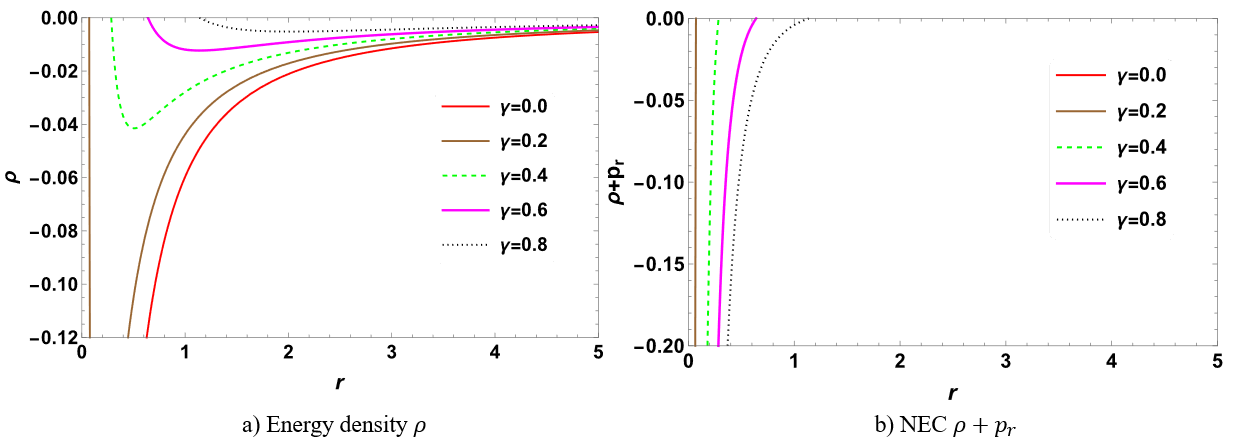}}
			{\includegraphics[scale=0.44]{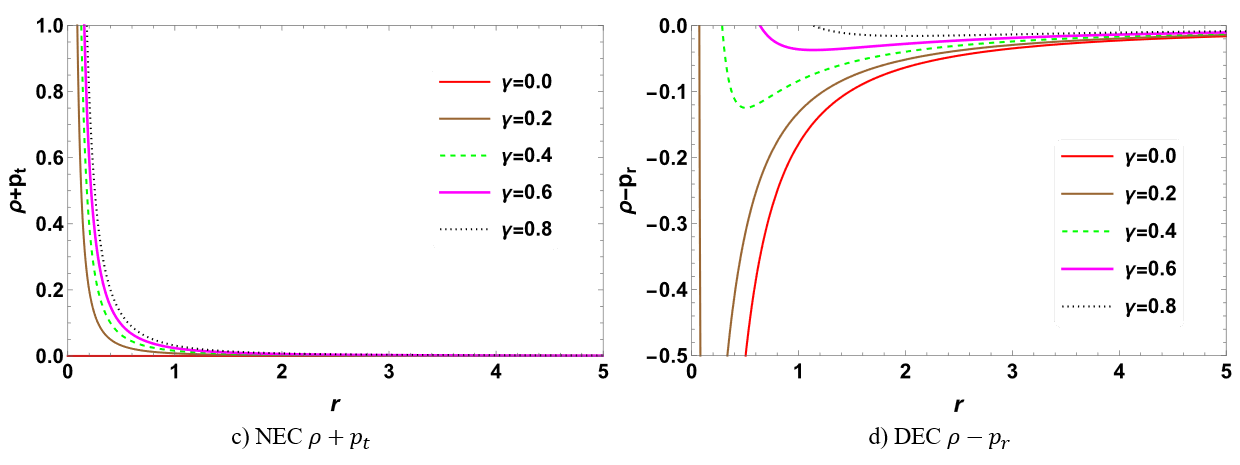}}
			{\includegraphics[scale=0.44]{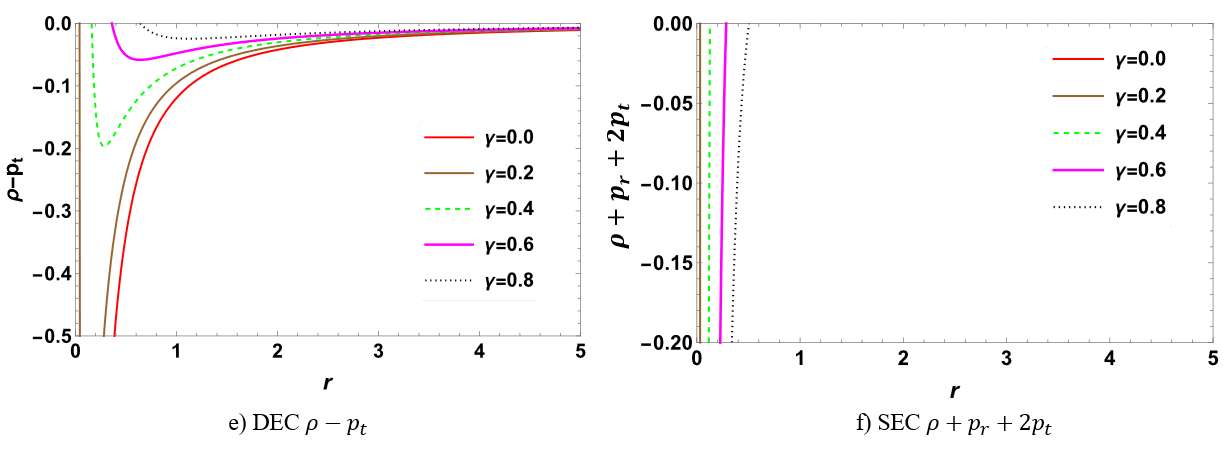}}
			\caption{\label{IMG12} Graphs illustrating the energy conditions radial density $\rho$, NEC, DEC, WEC, and SEC for non-baryonic matter, specifically for phantom fluid with parameters $\omega = -2, \gamma\geq0, K_1=1,$ and $A_3 = 1.$}
		\end{center}
	\end{figure}
	\begin{itemize}
		\item In the case of a non-baryonic fluid, the violation of the NEC occurs specifically when the equation of state parameter $\omega=-0.5$ (indicative of dark energy) and $\omega=-2$ (phantom fluid). This violation happens concerning the condition $\rho+p_r\leq0$ [see Figs. \ref{IMG11}b, \ref{IMG12}b], while it remains valid concerning $\rho+p_t$ [see Figs. \ref{IMG11}b, \ref{IMG12}b] \cite{Ksh}.
		
		\item When considering the WEC, for $\omega=-0.5$ (representing dark energy) and $\omega=-2$ (phantom fluid), the violation occurs concerning $\rho\leq0$ [see Figs. \ref{IMG11}a, \ref{IMG12}a], and $\rho+p_r\leq0$ [see Figs. \ref{IMG11}b, \ref{IMG12}b], while it remains valid concerning $\rho+p_t$ [see Figs. \ref{IMG11}c, \ref{IMG12}c].
		
		\item When examining the DEC, for $\omega=-0.5$ (related to dark energy), and $\omega=-2$ (phantom fluid), the violation occurs concerning $\rho-p_r\leq0$ [see Figs. \ref{IMG11}d, \ref{IMG12}d] and $\rho-p_t\leq0$ [see Figs. \ref{IMG11}e, \ref{IMG12}e].

		\item The SEC remains valid for $\omega=-0.5$ (Dark energy) when $\gamma\geq0$, as shown in Fig. (\ref{IMG11}f). In the case of $\omega=-2$ (phantom fluid), the SEC is violated concerning $\rho+p_r+2p_t\leq0$ when $\gamma\geq0.3$ as shown in Fig. (\ref{IMG12}f),

	\end{itemize}

\subsection{Effect of anisotropy}\label{subsec4.3}
	
In this part, we explore anisotropy to understand the traits of anisotropic pressure. Analyzing anisotropy is important as it helps uncover the internal structure of a relativistic wormhole setup. The degree of anisotropy within the wormhole can be assessed through a specific formula \cite{Rahaman1,Sharif,Shamir}.
\begin{equation}
	\Delta=p_t-p_r
\end{equation}

\begin{figure}[hptb]
	\begin{center}
		{\includegraphics[scale=0.5]{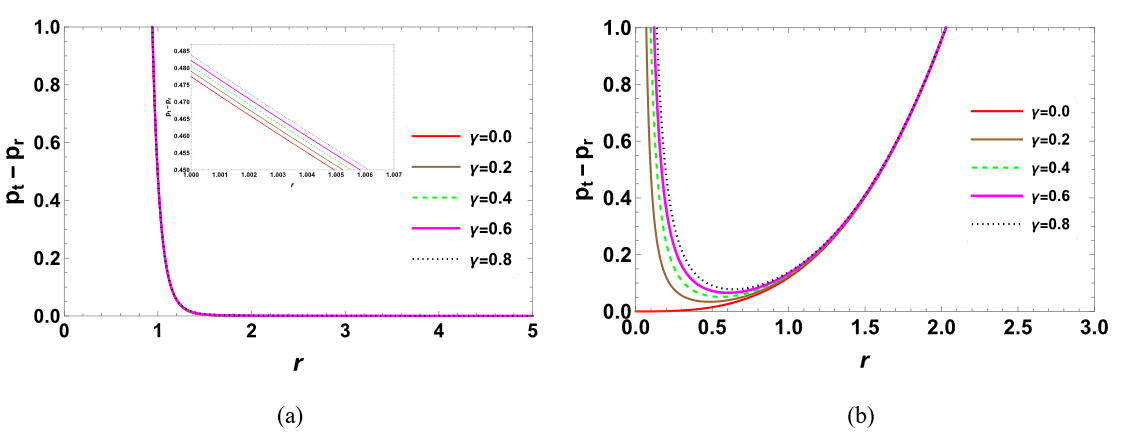}}
		{\includegraphics[scale=0.46]{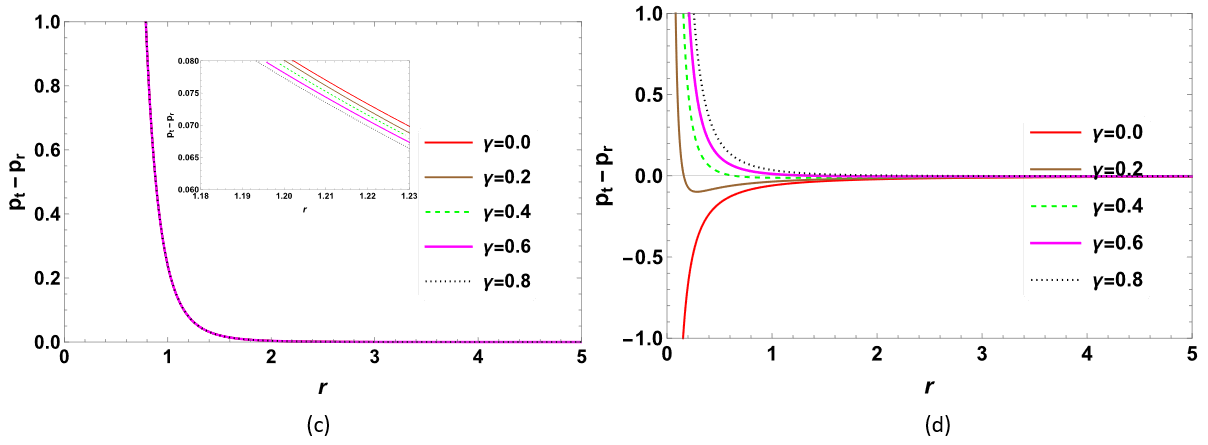}}
		\caption{\label{IMG13} Graphs illustrating the anisotropic parameter $\Delta=p_t-p_r$, (a) for baryonic fluid  with $\omega = \frac{1}{3}$, (b) for $\omega=1$. (c) For non-baryonic fluid with $\omega=-0.5$, (d) for $\omega=-2$,  with parameters  $\gamma\geq0, K_1=1,$ and $A_3 = 1.$}
	\end{center}
\end{figure}

 The parameter $\Delta$, serves as a measure of anisotropy known as the anisotropy factor in wormholes. When $\Delta$ is positive, the geometry of the wormholes is described as having a repulsive nature. Conversely, a negative $\Delta$ indicates an attractive geometry for the wormholes. If $\Delta =0$, it implies that the matter distribution within the wormholes possesses isotropic pressure, meaning the pressures in the radial and tangential directions are equal. The expression $2(p_t-p_r)/r$ represents the force attributed to an anisotropic matter source, known as the anisotropic force within wormholes. When the wormhole's geometry is considered attractive, this force acts inwardly. Conversely, in a scenario where the geometry is repulsive, the anisotropic force operates outwardly. In our analysis, for baryonic fluid, the anisotropy factor ($\Delta$) is shown as positive for the range  $\gamma\geq0$  [see Fig. \ref{IMG13}a, \ref{IMG13}b]. For non-baryonic fluid, the anisotropic factor $\Delta$ is positive for dark energy $\omega=-0.5$ shown in Fig. (\ref{IMG13}c) when $\gamma\geq0$. For $\omega=-2$, $\Delta$ is initially positive near the throat, its behavior changes to negative after crossing the throat, and it tends to zero for $\gamma\geq0.2$, see Fig. (\ref{IMG13}d).

	\section {Summary and conclusion}\label{sec.5}
	Wormholes are intriguing theoretical constructs within the realm of space-time. Despite their fascination, their existence in the physical world remains uncertain. In recent years, scientists have delved into the geometric properties of these hypothetical wormholes through mathematical investigations.
Here, we explored the possibility of a spherically symmetric Finslerian traversable wormhole solution undergoing conformal motion. We derived the field equations specific to this scenario using the barotropic linear EoS model. \\
	
	Some of the main features of the present investigation can be stated as follows:
	\begin{itemize}
		\item We extracted the conformal factor $\psi(r)$ and in Figs. (\ref{fig1bb}) and (\ref{fig1}) for baryonic fluid, and Figs. (\ref{fig2bb}) and (\ref{fig2cc}) non-baryonic fluid displays the profile of the derived shape function $(b(r)$, depicting its key characteristics. The figures illustrate the geometric configurations of a wormhole for different values of $\omega$ with a set of increasing Finslerian parameter $\gamma$, where $b(r)$ is a monotonically increasing function with $b(r) >0$, for all $r >r_0$.
		
		\item 	 Additionally, the wormhole maintains a finite proper radial distance function when $b(r) > 0$ for all $r > r_0$. The derivative $b'(r)$ remains less than 1, indicating compliance with the flaring-out condition. Furthermore, $b(r_0) - r_0 < 0$, establishing the throat radius as detailed in Table (\ref{tbl1}) and (\ref{tbl2}). The derived shape functions meet all necessary conditions, excluding asymptotic flatness, aligning with the findings of \cite{Mustafa2,Mustafa3}.
		
		\item The behavior of radial distance $l(r)$ various wormhole radius $r$ for different values of $\gamma$ in Fig. (\ref{fig1aa}) and (\ref{fig2aa}) from these two plots we observe that proper radial distance $l(r)$  remain stable throughout space-time.
		
		\item From Fig. (\ref{fig1a}) and (\ref{fig2a}), we observed the behavior of the radial velocity $V(r)$ after traversing the wormhole throat signifies its dynamic nature. An increase in velocity beyond the throat indicates an outward acceleration, suggesting possible expansion or movement away from the central region see Fig. (\ref{fig1a}) for baryonic fluid. Conversely, a decrease in velocity post-throat crossing implies an inward deceleration or potential collapse towards the core of the wormhole for non-baryonic fluid see Fig. (\ref{fig2a}).
		
		\item We discussed various energy conditions, including the NEC, WEC, DEC, and SEC, about the parameters $(\gamma, r)$. The violation of these energy conditions indicates the possible presence of exotic matter (either baryonic or non-baryonic). Exotic matter is a theoretical concept that possesses peculiar properties, such as negative energy densities or violations of the NEC or WEC, and plays a crucial role in the formation and maintenance of traversable wormholes. Exotic matter is postulated to generate the necessary negative energy to sustain the traversable wormhole, leading to intriguing possibilities for interstellar travel and potential connections between distant regions of the universe.
		
		\item 	We examined the effect of anisotropy described in Fig. (\ref{IMG13}) showing a physically plausible traversable wormhole model.  Within this context, it was observed that the geometric nature of the Finslerian wormhole exhibits a repulsive characteristic. It is essential to highlight that our study demonstrates that Finslerian wormholes, under conformal motion, within the barotropic linear equation of state, satisfy the criteria for being traversable wormholes. These findings suggest that Finslerian wormholes are physically plausible objects.
		
	\end{itemize}
	In our study, the parameter $\gamma$ in the Finslerian framework plays a significant role in the violation of energy conditions. When examining both baryonic and non-baryonic fluids, it is observed that all energy conditions are violated, which are specifically linked to the equation of state parameter $\omega$, when $\gamma \geq 0$. This violation suggests that the presence of either baryonic matter (such as radiation or stiff fluid) or non-baryonic matter (like dark energy or phantom fluid) near the wormhole's throat is evident under conformal transformation. Therefore, the present work leads to the conclusion that our newly formulated Finslerian wormhole becomes traversable as the conformal transformation results.
	The scope of the present investigating how conformal motion can help determine the properties of the photon sphere and the shadow of black holes in Finsler space-time by analyzing how conformal transformations affect the trajectories of light rays also has applications in observational astrophysics potentially, offering new ways to detect and study these exotic objects. Exploring the influence of conformal transformations on the quantum properties of space-time near Finslerian wormholes can advance the development of quantum theories of gravity.


\end{document}